# Singularity Protocol for Cross Chain AMM without Intermediate Tokens or Bridges

*Sumit Vohra*
Singularity, Singapore

*Abstract*— Automated Market Makers (AMMs) are decentralized exchange protocols that provide continuous access of token liquidity without the need of order books or traditional market makers. However this innovation has failed to scale when it comes to cross chain swaps. Modern day cross chain swaps employ double-sided AMMs which are not only inefficient in terms of liquidity fragmentation but also require an intermediate token which possesses inherent volatility risk as well as blockchain and bridging risk in case of wrapped tokens. This paper describes the inefficiencies of existing AMM invariants related to their mixed polynomial nature and derives a new class of AMMs which don't have bi-state dependency of the assets being swapped. The paper proposes a novel way of value transfer swaps using the described invariant that not only mitigates the need for bi-state dependency but also eliminates the need for intermediate tokens or bridging. We further show how the novel mechanism results in efficient cross chain swaps that have less gas requirement and no bridging risks associated with it. The technology promises to solve cross chain swaps across any permutation of L1, L2 & L3 chains.

## I. Introduction To Dex

To dive deeper, we need to understand how existing AMM based DEX protocols operate. A Decentralized Exchange (DEX) is a type of cryptocurrency trading platform that operates without a central intermediary. Unlike traditional exchanges, where transactions are facilitated by a centralized entity, DEXs enable users to trade directly with one another. This is achieved through blockchain technology, smart contracts, liquidity pools and most importantly, automated market making.

To have a functioning DEX protocol, liquidity is the most crucial factor, as it ensures that assets can be easily bought or sold without causing significant price fluctuations. In a DEX, liquidity is maintained through the following mechanisms:

**Liquidity Pools:** DEXs utilize liquidity pools to facilitate trading. Since these are decentralized pools, they can be created by any user willing to stake their tokens in exchange for interest/rewards. We will refer to such users as LPs or Liquidity Providers here onwards, these are users who deposit pairs of tokens into a smart contract to create token pools that are required for the DEX to work. Each pair consists of the token being traded (e.g. Token A) and another token (e.g. Token B) which acts as a counterbalance. This creates a reserve of both tokens that trading users can exchange across. This logic can further be extended to N-dimensional pools where multiple tokens can be added to a pool.

**Automated Market Makers (AMMs):** Liquidity pools are managed by Automated Market Maker algorithms. These algorithms automatically determine the price of tokens based on the ratio of the tokens available in the pool. As traders make transactions, the ratio changes, which in turn adjusts the token prices. One popular AMM formula used is the constant product formula, which ensures that the product of the quantities of the two tokens in the pool remains constant.

**Swapping Mechanism:** When a user wants to trade one token for another, they send their tokens to the smart contract governing the liquidity pool. The smart contract calculates the appropriate exchange rate based on the pool's current ratios. The user receives the desired token in exchange for the sent tokens.

**Incentives for Liquidity Providers:** To encourage users to supply tokens to the liquidity pools, DEXs reward Liquidity Providers with a portion of the trading fees collected from transactions. Additionally, some DEXs issue their native tokens as rewards to Liquidity Providers

**Arbitrage Opportunities:** Price discrepancies between the

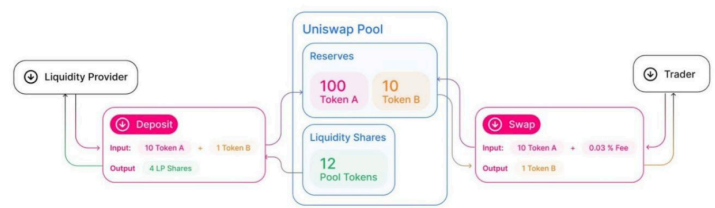

Figure1 : Uniswap pool management: htttps://docs.uniswap.org/contracts/v2/concepts/core-concepts/pools

liquidity pool and other exchanges create arbitrage opportunities. Bots can buy tokens from the DEX's liquidity pool at a lower price and sell them on another DEX/ platform at a higher price, thereby bringing the prices back into alignment.

By combining these mechanisms, DEXs aim to offer a trading environment where users can readily exchange assets while minimizing the impact of slippage and price volatility.

## II. Dex & Defi

A Decentralized Exchange (DEX) plays a pivotal role in the realm of Decentralized Finance (DeFi), which is a movement aimed at recreating and expanding upon traditional financial services using blockchain technology and decentralized networks. DEXs are a core component of the DeFi ecosystem due to their ability to provide secure, transparent, and permissionless trading of cryptocurrencies

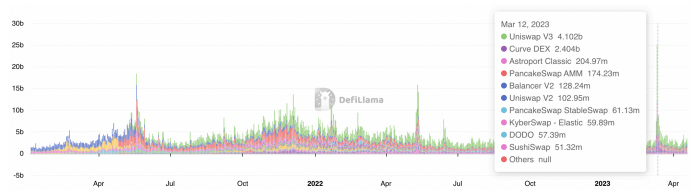

Figure2 : Total DEX transaction volume : https://defillama.com/

and tokens. Here's how DEXs fit into the broader landscape of DeFi:

**1. Eliminating Intermediaries:** One of the primary tenets of DeFi is the removal of intermediaries such as banks and financial institutions. DEXs align perfectly with this goal by allowing users to trade directly with each other, eliminating the need for a centralized exchange operator.

**2. Enhanced Security and Control:** DeFi emphasizes user control and ownership of assets. DEXs provide users with greater control over their funds since trades occur directly from their wallets. Users retain ownership of their private keys, reducing the risk of hacks and unauthorized access.

**3. Transparency and Audit-ability:** Transparency is a key feature of DeFi applications. DEXs leverage blockchain technology to record all transactions on a public ledger, enabling anyone to audit and verify trades. This transparency enhances trust within the DeFi ecosystem.

**4. Permissionless Access:** DEXs do not impose restrictions on who can participate in trading. As long as users have a compatible cryptocurrency wallet, they can access the DEX and start trading immediately. This permissionless nature aligns with DeFi's inclusive philosophy.

**5. Programmable and Composable Finance:** DeFi aims to democratize financial services by allowing users to program their financial interactions. DEXs integrate with DeFi protocols and smart contracts, enabling users to create sophisticated trading strategies, conduct automated trades, and execute complex financial operations.

**6. Liquidity Provision:** Liquidity is crucial for both DEXs and DeFi platforms. DEXs use liquidity pools to facilitate trading, and many DeFi protocols rely on these pools for operations like lending, borrowing, and yield farming. DEXs provide a marketplace for users to contribute liquidity and earn rewards.

In summary, DEXs are an essential component of the DeFi movement, providing users with a decentralized, transparent, and secure platform for trading assets directly. They contribute to the broader goals of DeFi by enabling permissionless access, user control, and programmable financial interactions within a decentralized ecosystem.

III. CURRENT AMM LANDSCAPE & ARCHITECTURES

A. *Major same chain AMM architectures*

1. *Uniswap*[1]*:* Uniswap is a decentralized exchange protocol built on Ethereum. It uses a simple Automated Market Maker (AMM) mechanism based on the constant product formula. Uniswap's smart contracts manage liquidity pools for different token pairs. It uses liquidity pools where users can deposit funds and receive pool tokens in return, which represent their share of the liquidity pool.

**Core Formula**: The Uniswap invariant is based on the constant product formula:

The Uniswap equation is given by:
$$x \cdot y = k$$
$x$ : amount of the first token in the pool,
$y$ : amount of the second token in the pool,
$k$ : constant product that remains unchanged during trades.

**Trading Mechanism**: When a trade is executed, the invariant ensures that the product of the token balances remains constant. As one token is bought, the other is sold, adjusting the balances to maintain the constant product. This results in slippage as the trade size increases.

**Key benefits and limitations**: Uniswap provides the basic AMM model to allow decentralized exchange of assets without the need for any centralized order book or parties. One key short fall of this technique is liquidity fragmentation, because each pool can only have 2 assets and hence the number of pools required grows at the rate of N x N, where N is the number of assets supported. Another short fall is that the standard AMM curve does not perform well for stable asset pools.

2. *Balancer*[2]*:* Balancer is a more complex AMM that allows users to create liquidity pools with multiple tokens and customizable weightings. It's designed to offer >= 2 size liquidity pools that reduces liquidity fragmentation and creates better trading strategies.

$$\prod_{i=1}^{n} \text{balance}[i]^{w[i]} = K$$

Where:
$n$ : number of tokens in the pool,
$\text{balance}[i]$ : balance of the $i$-th token in the pool,
$w[i]$ : weight assigned to the $i$-th token,
$K$ : constant product that remains unchanged during trades.

**Core Formula**: Balancer's invariant considers token weights and balances in the pool:

**Trading Mechanism**: When a trade is executed, the invariant ensures that the product of the token balances raised to their weights remains constant. As one token is bought, the other is sold, adjusting the balances to maintain the constant product.

**Key benefits and limitations**: Balancer improves on the Uniswap model by dramatically reducing liquidity fragmentation. However it still does not solve for stable tokens.

3. *Curve*[3]*:* Curve is optimized for stablecoin trading, aiming to minimize slippage by focusing on assets with similar values. It employs a bonding curve with a specialized formula.

**Core Formula**:

The expression is given by:

$$A \cdot n^n \cdot \sum_{i=1}^{n} X_i + D = AD \cdot n^n + \frac{D^{n+1}}{n^n \cdot \prod_{i=1}^{n} X_i}$$

Where:

$A$ : Amplification factor, a constant that adjusts sensitivity,
$n$ : Number of coins in the pool, indicating the number of tokens,
$D$ : Invariant, a constant value maintained across the swaps,
$X_i$ : Balance of the $i$-th token, representing the quantity of token $i$.

**Trading Mechanism**:

i. *Liquidity Pools Setup*: Curve operates through liquidity pools containing similar or pegged assets, such as different types of stablecoins. The pools are designed to maintain stable value ratios between the tokens, allowing traders to exchange assets with minimal slippage.

ii. *Virtual Balances and Price Model*: Curve introduces the concept of "virtual balances". Each token's balance is internally represented as a virtual balance to maintain stable value ratios. The price model of Curve is designed to minimize price slippage across different stable assets. This is achieved by focusing on stable value rather than the token's market price.

iii. *Amplification Factor* : Curve employs an "Amplification Factor" to adjust the sensitivity of the pool to the trading activities. Amplification factor allows traders to swap assets while maintaining stable value ratios. A higher amplification factor increases the pool's sensitivity to trades, allowing for more efficient swaps at the cost of a higher potential for impermanent loss.

Curve's trading mechanism aims to provide a stable and efficient trading experience for stablecoins and similar assets, minimizing price slippage while utilizing the amplification factor to adjust the pool's responsiveness to trading activities. This unique approach makes Curve particularly suitable for users seeking low-slippage trading in the stablecoin ecosystem.

B. *Cross Chain AMM Landscape*

The AMM examples shared above only work on swaps of tokens within the same chain. However, with the growing blockchain landscape and advent of L2 and L3 app-chains, token exchange across chains become a requirement rather than a feature.

Existing cross chain exchanges like Thorchain[4] or Axelar[5] employ double sided AMMs with a common token as a medium of exchange. So any exchange that happens using these protocols has to go through 2 AMM transactions (token1 → common_token ) & (common_token → token2) respectively.

Thorchain uses an intermediate token, called the "Rune token", and a relayer blockchain to support it. The swap involves moving Rune tokens internally within the relayer blockchain and executing 2 swaps on source and destination blockchain. This mechanism has inherent risks associated with operating a blockchain alongside bearing the burden of making sure that the price of the Rune token doesn't collapse or otherwise fluctuate wildly.

Axelar on the other hand follows a similar pattern but uses a stable token as the common token which is a bridged wrapped token corresponding to USDC called axlUSDC.

Though both of these exchanges enable cross chain swaps, they each have drawbacks associated with operating an underlying synthetic intermediary token as well an entire decentralised network. This also increases the trade cost in terms of maintaining the network as well executing double swaps.

This mechanism also limits the number of cross chain tradeable-blockchains as the network and the token needs to be extended to them.

Another method used by modern day L2s/L3s/appchains is to have localised DEXes based on wrapped tokens and employ bridges to generate those on their chains from their respective L1s. This has slightly reduced the trading cost but has increased the bridging risk.

To circumvent the above and natively bridge; protocols like CCTP have been launched but are heavily centralised and limited to just a couple of blockchains and only USDC which is non-extensible to everyday upcoming app-chains/app-tokens in the form of L2s and L3s.

C. *Why the standard same-chain AMM architectures cannot be applied in a cross chain environment*

AMMs have proven to be one of the most important innovation in DEX and DEFI protocols, with Uniswap being the largest DEX amongst all. Though the current variations of same-chain AMM architectures work very well for volatile and stable assets, they still fail to scale as cross-blockchain exchanges.

The primary reason for this is the structure of the invariant. All existing AMM invariants have a mixed polynomial nature meaning they require state consistency of both assets to be swapped. For e.g., for constant product market maker (Uniswap), we need pool balances of both x and y, which in the case of cross-chain swaps would be maintained as state variables on 2 different blockchains. Any calculation requiring both of them won't be able to achieve determinism or atomicity when executed across 2 different blockchain nodes. Also any change in the invariant because of liquidity injection or removal needs to be updated across the 2 blockchains atomically - which is impossible. This creates the need of heavy duty operations performed by cross chain AMM protocols as explained above. These operations though achieve the swap but come with huge liabilities in the form of bridging risk, liquidity fragmentation, maintaining synthetic tokens and balancing 2 sided AMMs.

We believe there is a more elegant solution to achieve highly scalable cross-chain swaps by extending the principles of same-chain AMM architectures and adapting them to a cross-chain world by fixing the invariant syncing issues that are introduced. This solution eliminates the risks and costs introduced by cross-chain AMM protocols used today.

## IV. EQUIVALENCE OF VALUE OF TRADE

To solve the above issues, and achieve cross chain swaps without the need of having above mentioned requirements, we propose a novel AMM protocol. The AMM protocol is designed to achieve cross-chain swaps without the need of state consistency. The algorithm revolves around value equivalence of trade. We state that for any swap to be optimal, the value of the tokens added to the pool should be equal to the value of tokens withdrawn from the pool. We use this property to create an AMM structure whose invariance is dependent only on local state variables available on each chain independently. In the next sections, we will define the properties of such liquidity pool and use that to derive swap and LP equations.

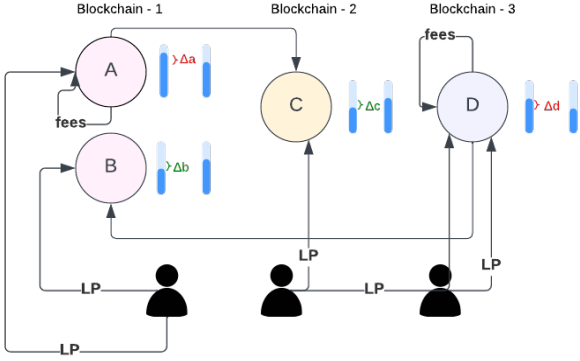

Figure 3 : AMM which enables swap to be executed cross chain. Shows swap between asset A and asset C | asset D and asset B. Liquidity can be provided across all blockchains. No bridge / no intermediate token / no intermediate blockchain required

### A. Properties of Liquidity Pool

We define the total value traded from the pool as 0. The mathematical equation for the same is given as:

$$\sum_{x=a}^{z} \int_{x_0}^{x_n} P(x)\,dx = 0$$

where: P(x) is the price function for each asset
assets in the pool range from {a to z} and balance of asset changes from initial balance $x_0$ to $x_n$.

In simple terms, this equation means that once the pool is setup with the initial deposit, the net change in value of the pool should always remain 0. We also have to make sure that the value change of an asset given by $\int_{x_0}^{x_p} P(x)\,dx$ can only happen on a swap transaction, i.e. any liquidity injection and removal have no effect on the same.
We need to ensure the above invariance holds in the following three cases:
    a) Swap two assets
    b) Inject/Remove Liquidity from trading pools
    c) Charging of trading fees

### B. Properties of Price Function Curve

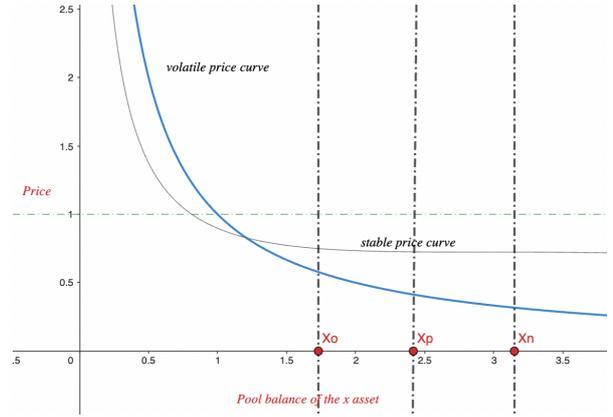

Figure 4: Price curve for volatile assets vs price curve for stable assets

Before we delve into these specific cases, we need to describe an ideal price function P(x) that would be negatively sloped in {0, ∞} to make sure that the price of the asset is inversely proportional to the quantity of the asset available in the pool. Also the price curve should have asymptotes at x axis=0 and y axis=0 to make sure that asset liquidity is available across all possible values which is a basic requirement for an AMM protocol. The price curve needs to be decreasing, differentiable and continuous for quantity>=0. A curve of such type would satisfy the following 4 properties:

$$\int_{x_0}^{x_p} P(x)\,dx + \int_{x_p}^{x_n} P(x)\,dx = \int_{x_0}^{x_n} P(x)\,dx$$

$$\frac{d}{dx}P(x) \leq 0 \quad \text{for } x \geq 0$$

$$P(0) \to \infty$$

$$P(\infty) \to 0$$

To derive such a curve, we can take motivation from Balancer's price equation denoted by $SP_{xv} = \frac{W_x \cdot v}{W_v \cdot x}$. This is spot price of asset $x$ wrt asset $v$.

A uni-variable price curve can be defined if we assume that there exists a virtual asset '$v$' which is never traded and has a constant balance. The above equation then converts to $SP_{xv} = C_v \cdot \left(\frac{W_x}{x}\right)$ where $C_v = \frac{v}{W_v}$
Since this would apply to all other assets in the pool, absorbing $C_v$ within $W_x$ such that $w_x = \{W_x\} \cdot C_v$ the price curve converts to $P_x = \left(\frac{W_x}{x}\right)$

Let's see if it satisfies the property of the ideal price curve and all 4 properties mentioned above.

**Satisfying 1:** $\int_{x_0}^{x_p} P(x)\,dx + \int_{x_p}^{x_t} P(x)\,dx = \int_{x_0}^{x_t} P(x)\,dx$

$$\int_{x_0}^{x_p} \frac{w_x}{x}\,dx + \int_{x_p}^{x_n} \frac{w_x}{x}\,dx = \int_{x_0}^{x_n} \frac{w_x}{x}\,dx \tag{0.1}$$

$$\int_{x_0}^{x_p} \frac{w_x}{x}\,dx = w_x \ln\left|\frac{x_p}{x_0}\right| \tag{0.2}$$

$$\int_{x_p}^{x_n} \frac{w_x}{x}\,dx = w_x \ln\left|\frac{x_n}{x_p}\right| \tag{0.3}$$

$$\int_{x_0}^{x_n} \frac{w_x}{x}\,dx = w_x \ln\left|\frac{x_n}{x_0}\right| \tag{0.4}$$

$$w_x \ln\left|\frac{x_p}{x_0}\right| + w_x \ln\left|\frac{x_n}{x_p}\right| = w_x \ln\left|\frac{x_n}{x_0}\right| \tag{0.5}$$

$$w_x \ln\left|\frac{x_n}{x_0}\right| = w_x \ln\left|\frac{x_n}{x_0}\right| \tag{0.6}$$

**Satisfying 2, 3 and 4:** These 3 properties are straightforward and we can see that they are easily met.

Having found our price function curve that satisfies the required properties, let's derive the swap equation that we need.

### C. Deriving Swap Value Equivalence Equation

We will use the pool invariant $\sum_{x=a}^{z} \int_{x_0}^{x_n} P(x)\,dx = 0$ to derive swap value equivalence equation. For a cross-chain swap between x=i and x=j, the swap equivalence is given by

$$\left(\int_{i_n}^{i_{n+1}} P_x\,dx\right) = -\left(\int_{j_n}^{j_{n+1}} P_x\,dx\right)$$

where $i_{n+1} = i_n + \Delta i$, $j_{n+1} = j_n - \Delta j$.

$$\sum_{x=a}^{z} \int_{x_0}^{x_n} P_x\,dx = 0 \tag{1.1}$$

$$\sum_{x=a}^{z} \int_{x_0}^{x_{n+1}} P_x\,dx = 0 \tag{1.2}$$

$$\sum_{x=a}^{z} \int_{x_0}^{x_n} P_x\,dx = \sum_{x=a}^{z} \int_{x_0}^{x_{n+1}} P_x\,dx \tag{1.3}$$

$$\sum_{x=i}^{j} \int_{x_0}^{x_n} P_x\,dx = \sum_{x=i}^{j} \int_{x_0}^{x_{n+1}} P_x\,dx \tag{1.4}$$

$$\left(\int_{i_0}^{i_n} P_x\,dx\right) + \left(\int_{j_0}^{j_n} P_x\,dx\right) = \left(\int_{i_0}^{i_{n+1}} P_x\,dx\right) + \left(\int_{j_0}^{j_{n+1}} P_x\,dx\right) \tag{1.5}$$

$$\left(\int_{i_n}^{i_{n+1}} P_x\,dx\right) = -\left(\int_{j_n}^{j_{n+1}} P_x\,dx\right) \tag{1.6}$$

Replacing with $P_x = \left(\frac{W_x}{x}\right)$:

$$\int_{x}^{i+\Delta i} \frac{W_x}{x}\,dx = -\int_{j}^{j-\Delta j} \frac{W_x}{x}\,dx \tag{2.1}$$

$$W_i [\ln(i+\Delta i) - \ln(i)] = -W_j [\ln(j-\Delta j) - \ln(j)] \tag{2.2}$$

$$\ln\left((i+\Delta i)^{W_i} \cdot (j-\Delta j)^{W_j}\right) = \ln\left((i)^{W_i} \cdot (j)^{W_j}\right) \tag{2.3}$$

$$(i+\Delta i)^{W_i} \cdot (j-\Delta j)^{W_j} = (i)^{W_i} \cdot (j)^{W_j} \tag{2.4}$$

$$\left(\frac{i+\Delta i}{i}\right)^{W_i} = \left(\frac{j}{j-\Delta j}\right)^{W_j} \tag{2.4}$$

The LHS of the equation 2.4 can be passed as a message using a relayer for inter-blockchain exchange and RHS can be calculated to find out $\Delta j$, the amount of token to be swapped out. Point to be noted here is that both LHS and RHS equations are uni-variate and only dependent on state variables of their own blockchain.

### D. Deriving Liquidity Provider / Fees Addition Equation

To sustain any AMM, Liquidity Providers are required. These LPs should be given a way to add and remove liquidity without disturbing the pool invariance and get rewarded in fees as incentive. This is in contrast to Uniswap/Balancer where addition/removal of liquidity changes the pool invariant.

We can't afford to have that in our pool invariant, because any changes in $\sum_{x=a}^{z} \int_{x_0}^{x_n} P(x)\,dx = K$, where K equals 0 in our case would require a state consistency that would not be possible to achieve in a cross blockchain environment.

In case of Singularity, LPs are given pool tokens in the form of *SINS* tokens which are proportionate to the value provided by the LP to the pool.

Initially when the pool is created, the total supply of SINS minted is set as a geometric mean of the number of tokens added to the pool which is similar to Uniswap v2.

$$\text{Initial Sinpool} = \sqrt[\text{number of elements}]{a_0 \cdot b_0 \cdot c_0 \cdot \ldots \cdot z_0}$$

After the initial liquidity, any additional liquidity should not disturb the $\int_{x_0}^{x_p} P(x)\,dx$ which corresponds to the value change of asset. This can be achieved via modifying the $x_0$ such that $\int_{x_0}^{x_p} P(x)\,dx = \int_{x_0 + \Delta x_0}^{x_p + \Delta x_p} P(x)\,dx$

$$\int_{x_0}^{x_p} P(x)\,dx = \int_{x_0+\Delta x_0}^{x_p+\Delta x_p} P(x)\,dx \tag{3.1}$$

$$w_x \ln(x_p) - w_x \ln(x_0) = w_x \ln(x_p + \Delta x_p) - w_x \ln(x_0 + \Delta x_0) \tag{3.2}$$

$$\ln(x_p) - \ln(x_0) = \ln(x_p + \Delta x_p) - \ln(x_0 + \Delta x_0) \tag{3.3}$$

$$\ln\left(\frac{x_p}{x_0}\right) = \ln\left(\frac{x_p + \Delta x_p}{x_0 + \Delta x_0}\right) \tag{3.4}$$

$$\frac{x_p}{x_0} = \frac{x_p + \Delta x_p}{x_0 + \Delta x_0} \tag{3.5}$$

$$x_p(x_0 + \Delta x_0) = x_0(x_p + \Delta x_p) \tag{3.6}$$

$$x_p \Delta x_0 = x_0 \Delta x_p \tag{3.7}$$

$$\Delta x_0 = \frac{x_0}{x_p} \Delta x_p \tag{3.8}$$

Where $x_0' = x_0 + \Delta x_0$ and $x_p' = x_p + \Delta x_p$ where $\Delta x_p$ corresponding to additional liquidity added to the pool in the form of fees or new liquidity.

After every trade, $x_0$ would be shifted as equation 3.8 to accommodate trade fees.
To accommodate liquidity addition and removal again equation 3.8 for $x_0$ adjustment and new *SINS* would be minted/burned $\text{new\_sins\_minted/burned} = \frac{\Delta x_p}{x_p} \cdot \text{sinpool}$

The above case is mentioned for all asset deposit, where all assets are deposited in ratio of their value in the pool.

We don't want to support single asset deposit in our v0 because of arbitrage opportunities it creates causing change in spot price. This leads to loss of value for the LP.

### E. Supporting Stable Assets Swaps

The above equations work well for pools with volatile assets, however we would need Stable Price curves for an efficient trade between stable assets. The optimal stable price curve needs to be constant across an asset equilibrium state given by $x_{stable}$ which is defined as the quantity of asset where $SP_{xy} = W$ where W is the predefined price ratio between assets. Let $Asset_x$ and $Asset_y$ be *USDC* and *DAI* respectively and W=1 then $x_{stable}$ is the quantity of USDC in the pool where $SP_{usdc/dai} = 1$ meaning $SP_{USDC} = SP_{DAI}$

We define the initial $x_{stable}$ and $y_{stable}$ as the $x_0$ and $y_0$ such that the initial quantities should reflect the correct ratio of the price.

We describe a stable price curve for above as
$$P_x = \frac{w}{x}\left(1 - \frac{A^2}{(x-x_{stable})^2 + A^2}\right) + \frac{w}{x_{stable}}\left(\frac{A^2}{(x-x_{stable})^2 + A^2}\right)$$

where A is the amplification factor. We use a modified version of '*Witch Of Agnesi*'[6] as the bell curve to flatten our price curve along $x_{stable}$. This bell curve is chosen to make P(x) integrable.

The above function satisfies satisfy the following 4 properties :

$$\int_{x_0}^{x_p} P(x)\,dx + \int_{x_p}^{x_n} P(x)\,dx = \int_{x_0}^{x_n} P(x)\,dx$$
$$\frac{d}{dx}P(x) \leq 0 \quad \text{for } x \geq 0$$
$$P(0) \to \infty$$
$$P(\infty) \to 0$$

Satisfying property no. 2 i.e. $\frac{d}{dx}P(x) \leq 0 \quad \text{for } x \geq 0$

$$P_x = \frac{w}{x}\left(1 - \frac{A^2}{(x-x_0)^2 + A^2}\right) + \frac{w}{x_0}\left(\frac{A^2}{(x-x_0)^2 + A^2}\right) \quad 4.1$$

$$\theta = \frac{A^2}{(x-x_0)^2 + A^2} \quad 4.2$$

$$\frac{d}{dx}(\theta) = -\frac{2A^2(x-x_0)}{(A^2 + (x-x_0)^2)^2} \quad 4.3$$

$$\frac{d}{dx}(P_x) = w\left(-\frac{1}{x^2} - \theta \cdot \frac{d}{dx}\left(\frac{1}{x}\right) + \frac{d}{dx}(\theta) \cdot \left(\frac{1}{x_0} - \frac{1}{x}\right)\right) \quad 4.4$$

**Observation for both** $x_0 > x$ **and** $x_0 < x$

The expression is $\leq 0$.

Satisfying 1, 3 and 4: These 3 properties are straightforward and we can see that they are easily met.
Having found our price function curve that satisfies the required properties, let's derive the swap equation that we need.

**Deriving Swap Value Equivalence Equation**
We will use the pool invariant $\sum_{x=a}^{z}\int_{x_0}^{x_n} P(x)\,dx = 0$ to derive swap value equivalence equation. For a trade between *i* and *j*, the swap equivalence is give by

$$\left(\int_{i_n}^{i_{n+1}} P_x\,dx\right) = -\left(\int_{j_n}^{j_{n+1}} P_x\,dx\right) \text{ where } i_{n+1} = i_n + \Delta i, \quad j_{n+1} = j_n - \Delta j.$$

Integral of the above P(x) is calculated to be

$$\left(w_x \ln(x) - \frac{w_x A}{x_{stable}} \cdot \arctan\left(\frac{x_{stable}-x}{A}\right) + \frac{w_x A\left(-2A\ln(x) + 2x_{stable}\arctan\left(\frac{x_{stable}-x}{A}\right) + A\ln(A^2 + (x_{stable}-x)^2)\right)}{2x_{stable}^2 + 2A^2}\right)$$

Let's call this function *I(x)*. We calculate the change in value as $I(i_n + \delta_i) - I(i_n)$ and pass this as a message for inter blockchain exchange. We further calculate the value of $I(j_n)$ and find $\delta_j$ using binary search as $I(j_n - \delta_j)$ is an increasing function. To reduce the gas, we can compute this off chain and just verify the function on chain.

**Changing Stable Quantities**
For the price equation to be univariate and at the same time the curve to have stable curve properties, the $x_{stable}$ needs to be changed only at the time of liquidity addition and removal. As an example let USDC and DAI initial quantities be 100 and 100, then addition of liquidity 200 USDC and 200 DAI results the $x_{stable}$ and $y_{stable}$ to change from 100 to 300.

The exact algorithm of movement will be described at the time of implementation and is based on simulations.

## V. CONCLUSION

In this paper, we presented a novel AMM architecture to swap between assets cross chain without the need of a bridge or a synthetic token and an intermediate blockchain. The proposed architecture not only reduces the security burden for achieving cross chain swaps but encourages much more efficient utilisation of liquidity.